\documentclass[aps,prd,twocolumn,amsmath,amssymb,amsfonts,nofootinbib,superscriptaddress,altaffilletter,showpacs]{revtex4}
\usepackage{amsgen,amstext,amsbsy,amsopn,amsfonts,amssymb}
\usepackage{supertabular}
\usepackage[pdftex]{graphicx}
\usepackage{hyperref}
\usepackage{graphpap}
\usepackage{rotating}
\usepackage{url}
\usepackage{color}
\usepackage{dcolumn}

\def\Var{\textrm{Var\,}}

\def\sci#1#2{{#1}{\times 10^{#2}}}

\def\SNR{{\textrm{SNR}}}

\def\Fshift{{{\Xi}}}

\def\hx{h_{\times}}
\def\hp{h_+}

\def\Ap{A_+}
\def\Ax{A_{\times}}

\def\Fp{F_+}
\def\Fx{F_{\times}}

\def\Re{\textrm{Re}}
\def\Im{\textrm{Im}}

\definecolor{green4}{rgb}{0,0.25,0}

\def\dochecks{0}
\def\check#1{{\if \dochecks 1 {\color{green4}\ \\ \rule{0.4\linewidth}{1pt}\hfill Check \hfill\rule{0.4\linewidth}{1pt} \hspace*{0pt} \par #1 \rule{\linewidth}{1pt} } \fi}}

\begin{document}

\title{Loosely coherent searches for sets of well-modeled signals.}
\author{V.~Dergachev}
\date{\today}
\address{
LIGO Laboratory,
California Institute of Technology,
MS 100-36,
Pasadena, CA 91125, USA
}
\begin{abstract}
We introduce a high-performance implementation of a loosely coherent statistic sensitive to signals spanning a finite-dimensional manifold in parameter space. Results from full scale simulations on Gaussian noise are discussed, as well as implications for future searches for continuous gravitational waves. We demonstrate an improvement of more than an order of magnitude in analysis speed over previously available algorithms. As searches for continuous gravitational waves are computationally limited, the large speedup results in gain in sensitivity.
\end{abstract}
\maketitle


\section{Introduction}

Loosely coherent algorithms~\cite{loosely_coherent} detect families of noise-dominated signals. The immediate application is the search for continuous gravitational waves. The large parameter space volume to be searched and the data volumes necessary to search for exceedingly weak signlas make our searches computationally limited. Hence improvements in efficiency of our algorithms directly affect the sensitivity of our results. As we are still waiting for the first detection, we cannot rely on a natural source to verify correctness of the detector and search pipelines. Our algorithms must be robust to possible imperfections of the detector, to faults in understanding of gravitation or even to bugs in the search programs. It also helps to be sensitive to a wide family of signals, in case the loudest source is not a perfect sine wave, which can result, for example, from a companion object.

The first implementation of a loosely coherent search \cite{loosely_coherent} was designed for signals with a large amount of phase deviation over 30-minute interval from a perfect sine wave. This provided the gain in sensitivity needed for follow up of outliers seen
in the full dataset of the LIGO detector's fifth science run S5~\cite{FullS5Semicoherent}, while preserving the robustness of 
the underlying, semi-coherent PowerFlux algorithm \cite{S4IncoherentPaper, PowerFluxTechNote, PowerFlux2TechNote}.

In this paper we explore the other end of the spectrum - an algorithm sensitive to coherent signals described by a small number of parameters, such as frequency or sky position. A number of coherent codes have been developed previously, in particular \cite{jks, BCCS, BC00, BayesianFstat, resampling, Astone1, Astone2, Krolak1, Curt1, Curt2}. What is different in our approach is that, unlike previous algorithms, our sky templates are ``thick'', sweeping small patches of parameter spaces. In particular, individual signals taken from the middle of nearby patches do not have a high degree of overlap. This property, together with careful attention to implementation particulars, provides for a very high performance coherent code.

\section{\label{coherence_engine}A simplified algorithm}
Suppose that we are interested in a family of signals described by the formula 
\begin{equation}
\label{simplified_signal}
s(t, a)=Ae^{2\pi i\nu \left(t+ \Fshift(t, a)\right)}
\end{equation}
where $\nu$ is the signal frequency and $\Fshift(t, a)$ has a smooth dependence on time $t$ and multidimensional parameter set $a$.

Our input data $f(t)=s(t, a)+\xi(t)$ consists of one signal from this family and, ideally, uncorrelated Gaussian noise of variance $\sigma^2(t)$:
\begin{equation} 
\left\langle \xi(t)\xi(t') \right\rangle=\sigma^2(t)\delta(t-t')
\end{equation}

If we knew the frequency $\nu$ and parameter set $a$, we could form a matched filter that would return the amplitude of our signal:
\begin{equation}
A(\nu, a)=\frac{1}{\mathfrak W}\int_{t_0}^{t_1} \frac{f(t)}{\sigma^2(t)} e^{-2\pi i\nu\left(t+\Fshift(t, a)\right)}dt
\end{equation}
Here $\mathfrak W$ is the total weight:
$$
\mathfrak W=\int_{t_0}^{t_1} \frac{1}{\sigma^2(t)} dt
$$

If the signal parameters are not known, one can construct a bank of waveforms $s(t)$ and evaluate the integral separately for each template. This is, of course, computationally expensive. 

One way to gain a large speedup is to introduce a new time variable $t'$ that straightens out our signal into a sine wave:
\begin{equation}
t'=t+\Fshift(t, a)
\end{equation}
One then resamples \cite{resampling} the input data $f(t)$ to be equally spaced in the new variable $t'$ and uses a Fourier transform to compute amplitudes for a range of possible signal frequencies. It has also been proposed by B.F.Schutz that a change to nearby value $a'$ can be accomplished with a kernel \cite{kernel}. As far as we know this method of ``stepping around in the sky'' does not yet have an implementation.

A simpler approach that combines the best features of both resampling and stepping around in the sky is to consider the following function of the input data:
\begin{equation}
F(\lambda ; \nu_0, a_0)=\frac{1}{\mathfrak W}\int^{t_1}_{t_0} \frac{f(t)}{\sigma^2(t)} e^{-2\pi i\nu_0(t+\Fshift(t,a_0))}e^{-2\pi i\lambda t} dt
\end{equation}
which is easily computed with a fast Fourier transform. For $\lambda=0$ it returns an amplitude estimate of the signal with parameters $(\nu_0, a_0)$. The values of $F$ for $\lambda\neq 0$
 carry slightly distorted information on nearby templates, which can be used to compute estimates of their signal amplitude with a convolution:
\begin{equation}
\begin{array}{l}
A(\nu, a)=\displaystyle \frac{1}{\mathfrak W}\int_{t_0}^{t_1} f(t) e^{-2\pi i\nu\left(t+\Fshift(t, a)\right)}\frac{1}{\sigma^2(t)}dt = \\
\quad	=\displaystyle\frac{1}{\mathfrak W}\int_{t_0}^{t_1} f(t) e^{-2\pi i\nu_0\left(t+\Fshift(t, a_0)\right)}\frac{1}{\sigma^2(t)} \cdot \\
\quad	 \quad  \cdot e^{-2\pi i(\nu-\nu_0)t-2\pi i(\nu-\nu_0)\Fshift(t, a_0) -2\pi i \nu(\Fshift(t, a)-\Fshift(t, a_0))} dt = \\
\quad	=\displaystyle\int   F(\nu-\nu_0-\mu ; \nu_0, a_0) \cdot \\
\quad	 \quad \cdot \displaystyle\int_{t_0}^{t_1} e^{-2\pi i\mu t-2\pi i(\nu-\nu_0)\Fshift(t, a_0) -2\pi i \nu(\Fshift(t, a)-\Fshift(t, a_0))} dt d\mu 
\end{array}
\end{equation}

The reader will notice that the last term is not quite the ordinary convolution - one of the convolved terms has a (slow) dependence on the convolution parameter. We call this a ``pseudo-convolution'' operator. We distinguish this case from the more general notion of integral operator, because in practical computation we do not have to update the slowly changing convolution with every data sample and the computational requirements of the operator are equivalent to the computational requirements of plain convolution.

In our case the pseudo-convolutions are of the form
\begin{equation}
A(\nu, a)=\int   F(\nu-\nu_0-\mu ; \nu_0, a_0) 
\int_{t_0}^{t_1} e^{-2\pi i\mu t-U(t, \nu, a)} 
 dt d\mu
\end{equation}
where $U(t, \nu, a)$ is a {\em phase mismatch} function describing difference in phase evolution between nearby templates. For smooth $U(t, \nu, a)$ the convolution operator is close to $\delta$-function. In practical computation, using discrete Fourier transform, this means that our convolution can be approximated with an FIR filter that has small number of terms.

It is crucial to control the number of terms in the convolutions. There are two ways to achieve that, aside from simply using small increments in the parameters with a corresponding increase in the number of templates.

First, we can subtract a linear term from the argument of the exponent so that it has the same value at both ends of the segment $[t_0, t_1]$. The linear term is analogous to a Doppler shift correction and results in relabeling of frequency parameter $\nu$.

Secondly, pseudo-convolutions have small or null commutators. One can then apply methods of linear algebra to change from initial set of operators (usually corresponding to individual parameters) to a set with progressively fewer convolution terms. The operators with the smallest number of terms are used in the innermost computational loop thus determining overall performance of the code.

We implement these techniques by representing $U(t, \nu, a)$ as a sum of a linear term and Fourier series with coefficients linear in $\nu$:
\begin{equation}
\begin{array}{l}\displaystyle
U(t, \nu, a)\simeq \frac{U(t_1, \nu, a)-U(t_0, \nu, a)}{t_1-t_0}\left(t-t_0\right)+\\
\qquad\qquad\qquad\displaystyle +\sum_{k=-\infty}^{\infty} \left(u_k(a)^0+u_k^1(a)\nu\right)e^{2\pi i k t }
\end{array}
\end{equation}
In most cases the equality is exact and only one of $u_k^0$ or $u_k^1$ is non-zero. The set of pseudo-convolution operators can be transformed into a more convenient basis by minimizing coefficients in the series, which is then converted to its Fourier transform with the help of the Jacobi-Anger identity:
\begin{equation}
e^{iz\cos \theta}=\sum_{n=-\infty}^{\infty} i^n J_n(z)e^{in\theta}
\end{equation}
Only a small number of terms are usually needed and recomputation is done rarely. The simulations presented in section \ref{sec:simulations} were carried out with innermost loops that used convolutions with only 11 terms - a number chosen to take advantage of vectorized arithmetic on modern CPUs.

\section{Implementation details}

While a fast engine to compute coherent sums is essential for our search code, it is only part of a whole. In particular, after computing coherent sums one needs to derive statistics such as maximum SNR or upper limit which can be expensive to compute. For example, if one were to use a rank-based method (which is nicely robust), it would require sorting the computed sum which has a scaling of $N\log N$ - same as a fast Fourier transform, and much slower than a convolution.

It is usually not practical to analyze the entire band of interest in one go, but rather one splits it into frequency bands of $1$~Hz or smaller. The amount of loaded data can be greatly reduced by precomputing short discrete Fourier transforms (SFTs) of duration commensurate with the region of interest. It is convenient to have the SFT length be short enough that the signal frequency can be assumed to be stationary.

\subsection{Polarization analysis}
Continuous gravitational waves have a more complicated form than is given by equation \ref{simplified_signal} - there are two polarizations 
with detected strengths that vary with the orientation of the detector.

The following analysis is similar to one found in \cite{jks, FstatTechNote, PowerFluxPolarizationNote}; we prefer, however, to reduce the four real parameters to two complex parameters that have a symmetric role. We also derive a convenient equation for surfaces of constant $h_0$.

We start by assuming that our signal consists of two polarizations:
\begin{equation}
\begin{array}{rcl}
\hp'&=&\Ap\cos(\omega t+\phi) \\
\hx'&=&\Ax\sin(\omega t+\phi) \\
\end{array}
\end{equation}

A generic pulsar signal can be represented as $\Ap=h_0\left(1+\cos^2(\iota)\right)/2$, 
$\Ax=h_0\cos(\iota)$, with $h_0=\Ap+\sqrt{\Ap^2-\Ax^2}$ and 
$\cos(\iota)=\Ax/\left(\Ap+\sqrt{\Ap^2-\Ax^2}\right)$

We will assume that demodulation is performed for a fixed frame of plus and cross polarizations rotated at an angle $\beta$. 
In this coordinate system we have:
\begin{equation}
\begin{array}{rcl}
\hp&=&\Ap\cos(\omega t+\phi)\cos(\epsilon)-\Ax\sin(\omega t+\phi)\sin(\epsilon) \\
\hx&=&\Ap\cos(\omega t+\phi)\sin(\epsilon)+\Ax\sin(\omega t+\phi)\cos(\epsilon) \\
\end{array}
\end{equation}
where we introduced $\epsilon=2(\psi-\beta)$.

The signal amplitude in SFT bin corresponding to frequency $\omega$ is then
\begin{equation}
\begin{array}{rcl}
z&=&\displaystyle\int \left(\Fp\hp+\Fx\hx\right) e^{-i\omega t} dt=\\
 &=&\vphantom{\int}\frac{1}{2}e^{i\phi}\left(\Fp(\Ap \cos(\epsilon)+i\Ax \sin(\epsilon))+\right.\\
 &&\qquad\left.+\Fx(\Ap \sin(\epsilon)-i\Ax \cos(\epsilon))\right)\\
&=&\Fp w_1+\Fx w_2
\end{array}
\end{equation}
where we introduced complex amplitude parameters
\begin{equation}
\begin{array}{rcl}
w_1&=&\frac{1}{2}e^{i\phi}(\Ap \cos(\epsilon)+i\Ax \sin(\epsilon))\\
w_2&=&\frac{1}{2}e^{i\phi}(\Ap \sin(\epsilon)-i\Ax \cos(\epsilon))\\
\end{array}
\end{equation}
The complex amplitude parameters $w_1$ and $w_2$ are algebraically symmetric:
\begin{equation}
\begin{array}{rcl}
b&=&\displaystyle|w_1|^2+|w_2|^2=\frac{1}{4}(\Ap^2+\Ax^2)=\\
 &&\qquad=\displaystyle\frac{1}{16}h_0^2(1+6\cos^2(\iota)+\cos^4(\iota)) \\
a&=&\displaystyle\Im (w_1\bar{w}_2)=\frac{1}{4}\Ap\Ax=\frac{1}{8}h_0^2(1+\cos^2(\iota))\cos(\iota)
\end{array}
\end{equation}
and are otherwise unconstrained.
They have a simple relation to more familiar real amplitude parameters $\mathcal A^\mu$ \cite{mock_lisa}:
\begin{equation}
\begin{array}{rcl}
w_1&=&{\mathcal A}^1-i{\mathcal A}^3\\
w_2&=&{\mathcal A}^2-i{\mathcal A}^4\\
\end{array}
\end{equation}
One easily finds the following equation of constant $h_0$:

\begin{equation}
\sqrt{|w_1+iw_2|}+\sqrt{|w_1-iw_2|}=\sqrt{h_0}
\end{equation}
the solutions of which form a singular surface enclosing a non-convex solid. This complicated form is responsible for differences between PowerFlux style upper limits, which are always limited by sensitivity to linearly polarized signals, and SNR statistics, the outliers of which can have arbitrary polarization. A related issue is the difference between the $\mathcal F$-statistic and the $\mathcal B$-statistic \cite{BayesianFstat}.


It is easy to compute generators for rotations in $\phi$ and $\epsilon$:
\begin{equation}
\frac{\partial}{\partial \phi}\left(\begin{array}{c}
                                    w_1 \\
				    w_2
                                    \end{array}\right)=i\left(\begin{array}{c}
                                    w_1 \\
				    w_2
                                    \end{array}\right)
\end{equation}
\begin{equation}
\frac{\partial}{\partial \epsilon}\left(\begin{array}{c}
                                    w_1 \\
				    w_2
                                    \end{array}\right)=\left(\begin{array}{cc}
                                    0 & -1 \\
				    1 & 0
                                    \end{array}\right)\left(\begin{array}{c}
                                    w_1 \\
				    w_2
                                    \end{array}\right)
\end{equation}
This shows that the surface of $h_0=1$ is obtained by revolving the parabola
\begin{equation}
\begin{array}{rcl}
w'_1 &=& \frac{1}{4}(1+\cos^2(\iota)) \\
w'_2 &=& -\frac{i}{2} \cos(\iota)\\
\end{array}
\end{equation}
along $\phi$ and $\epsilon$.

\subsection{Coherent sum}
In a coherent analysis we have data for many SFTs $\left\{z_i\right\}_{i=1}^N$ which contain our signal mixed with instrumental noise
\begin{equation}
z_i=\left(\Fp(t_i) w_1+\Fx(t_i) w_2\right)e^{ i \Phi(t_i)}+\xi_i
\end{equation}
and we construct a weighted sum
\begin{equation}
Z=\sum_{i=1}^{N} \alpha_i \frac{z_i e^{ -i \Phi(t_i)}}{\Fp(t_i) w_1'+\Fx(t_i) w_2'}
\end{equation}
which estimates signal amplitude. Here $\Phi(t_i)$ describes some assumed phase evolution due to changes in the source or detector, $\alpha_i$ are weights satisfying $\sum_{i=1}^N \alpha_i=1$, and $w_1'$ and $w_2'$ are computed for polarization and phase of our signal, but assuming $h_0=1$, i.e. they satisfy
\begin{equation}
\label{eqn:wh0const}
\sqrt{|w_1'+iw_2'|}+\sqrt{|w_1'-iw_2'|}=1
\end{equation}

There are many ways to compute ``optimal'' weights $\alpha_i$, in particular \cite{BayesianFstat, jks}. Here we use the variance of $Z$ as the optimality measure. Assuming $\xi_i$ are independent Gaussian variables with zero mean, we compute:
\begin{equation}
\Var(Z)=\sum_{i=1}^{N} \alpha_i^2 \frac{\Var(\xi_i)}{|\Fp(t_i) w_1'+\Fx(t_i) w_2'|^2}
\end{equation}
One easily finds that $\Var(Z)$ is minimized for
\begin{equation}
\alpha_i=\frac{1}{\mathfrak A(w_1', w_2')}\frac{|\Fp(t_i) w_1'+\Fx(t_i) w_2'|^2}{\Var(\xi_i)}
\end{equation}
where $\mathfrak A$ is the normalization weight:
\begin{equation}
\mathfrak A(w_1', w_2')=\sum_{i=1}^N \frac{|\Fp(t_i) w_1'+\Fx(t_i) w_2'|^2}{\Var(\xi_i)}
\end{equation}
Substituting the optimal coefficients $\alpha_i$, we compute:
\begin{equation}
\label{eqn:z}
Z(w_1', w_2')=\frac{1}{\mathfrak A(w_1', w_2')}\sum_{i=1}^{N} z_ie^{ -i \Phi(t_i)} \frac{\Fp(t_i) \bar{w}_1'+\Fx(t_i) \bar{w}_2'}{\Var(\xi_i)}
\end{equation}
and
\begin{equation}
\label{eqn:var_z}
\Var(Z(w_1', w_2'))=\frac{1}{\mathfrak A(w_1', w_2')}
\end{equation}
We see that the total weight $\mathfrak A(w_1', w_2')$ can be interpreted as a measure of the amount of data used to compute $Z(w_1', w_2')$, as in the case of stationary input data it is proportional to the number of independent SFTs.

We now note that $\mathfrak A(w_1', w_2')$ depends quadratically on coefficients $w_i$ and on the estimate of variance of the data and can thus be computed once for all templates with similar detector response $\Fp$ and $\Fx$. The unnormalized coherent sum ${\mathfrak A} Z$ is linear in $w_i$, and these coefficients are exactly the kind of sum that we learned to compute in section \ref{coherence_engine}.

Our computation then results in the following representation of $Z$:
\begin{equation}
Z(n)=\frac{ X_1(n)\bar{w}_1'+X_2(n)\bar{w}_2'}{Y_{11} |w_1'|^2+2Y_{12} \Re(w_1'\bar{w_2}')+Y_{22}|w_2'|^2}
\end{equation}
Here $Y_{11}$, $Y_{12}$ and $Y_{22}$ are coefficients that determine the total weight of the coherent sum and $X_1(n)$ and $X_2(n)$ are two arrays (in frequency bin $n$) of coherent sums:

\begin{equation}
\label{eqn:sum_coeffs}
\begin{array}{lcl}
Y_{11}&=& \displaystyle\sum_{i=1}^N \frac{\Fp(t_i)^2}{\Var(\xi_i)}\\ 
Y_{12}&=& \displaystyle\sum_{i=1}^N \frac{\Fp(t_i)\Fx(t_i)}{\Var(\xi_i)}\\ 
Y_{22}&=& \displaystyle\sum_{i=1}^N \frac{\Fx(t_i)^2}{\Var(\xi_i)}\\ 
X_1&=& \displaystyle\sum_{i=1}^{N} z_ie^{ -i \Phi(t_i)} \frac{\Fp(t_i)}{\Var(\xi_i)}\\ 
X_2&=& \displaystyle\sum_{i=1}^{N} z_ie^{ -i \Phi(t_i)} \frac{\Fx(t_i)}{\Var(\xi_i)}\\ 
\end{array}
\end{equation}

For convenience, we tabulate a few useful expressions using these coefficients:

\begin{equation}
\label{eqn:defs1}
\begin{array}{lcl}
\mathfrak{A}(w_1', w_2')&=&Y_{11} |w_1'|^2+2Y_{12} \Re(w_1'\bar{w_2}')+Y_{22}|w_2'|^2\\
Z(w_1', w_2')&=&\displaystyle\frac{X_1\bar{w}_1'+X_2\bar{w}_2'}{\mathfrak{A}(w_1', w_2')} \\
\textrm{SNR}(w_1', w_2')&=& \displaystyle\frac{\left|X_1\bar{w}_1'+X_2\bar{w}_2'\right|^2}{\mathfrak{A}(w_1', w_2')} \\
\textrm{SNR}_R(w_1', w_2')&=&\displaystyle \frac{\left(\Re\left(X_1\bar{w}_1'+X_2\bar{w}_2'\right)\right)^2}{\mathfrak{A}(w_1', w_2')} \\
\end{array}
\end{equation}

\subsection{Efficient computation of coherent sum statistics}
We now need to reduce the data to the SNR or some other statistic of the loudest outlier. The simplest method, employed by PowerFlux \cite{S4IncoherentPaper, PowerFluxTechNote, PowerFlux2TechNote} and the large-$\delta$ loosely coherent search \cite{loosely_coherent} is to scan different values of $w_1$ and $w_2$ looking for a maximum. The elements of our coherent sums are computed, however, with $\sim 22$ complex multiplications for each frequency bin, and even a modest grid of polarization parameters dominates computation.

An approach taken in \cite{jks} is to analytically maximize SNR over $w_1$ and $w_2$. A new $\mathfrak B$-statistic was introduced in \cite{BayesianFstat} that was shown to have a better physically motivated prior.

\begin{table*}
\begin{tabular}{ll}
Statistic & Formula \\
\hline \\
Raw power & $\left|X_1\right|^2+\left|X_2\right|^2$\\
Adjusted power & $Y_{22}\left|X_1\right|^2+Y_{11}\left|X_2\right|^2$\\
SNR & $\displaystyle\max_{w_1', w_2'} \frac{\left|X_1\bar{w}_1'+X_2\bar{w}_2'\right|^2}{\mathfrak A(w_1', w_2')}$ \\
Upper limit & $\displaystyle\max_{w_1', w_2'} \sqrt
	{\frac{\left|X_1\bar{w}_1'+X_2\bar{w}_2'\right|^2}{\mathfrak A(w_1', w_2')^2}+
	2\frac{\left|X_1\bar{w}_1'+X_2\bar{w}_2'\right|}{\mathfrak A(w_1', w_2')^{3/2}}+
	\frac{\sigma-1}{\mathfrak A(w_1', w_2')}
	}$\\
$\mathfrak F$-stat& $\displaystyle {\frac{Y_{22}|X_1|^2-2 Y_{12} \Re(X_1\bar{X_2})+Y_{11}|X_2|^2}{Y_{11}Y_{22}-Y_{12}^2}}$ \\
$\mathfrak B$-stat& $\displaystyle \int \textbf{dw}' \frac{1}{{ \mathfrak A(w_1', w_2')}}\Theta\left(\frac{\left|X_1\bar{w}_1'+X_2\bar{w}_2'\right|^2}{2 \mathfrak A(w_1', w_2')}\right)$\\
\end{tabular}
\caption{Statistics functions. The variables $w_1'$ and $w_2'$ are constrained by equation \ref{eqn:wh0const}.} 
\label{tab:statistics}
\end{table*}

\check{
Derivation of $\mathcal F$-statistic:
\begin{equation}
\mathcal{L}(x, \mathcal{A})=e^{\mathcal{A}^\mu x_\mu-\frac{1}{2}\mathcal{A}^\mu M_{\mu\nu} \mathcal{A}^\nu} =e^{\Re(\bar{w}_1 X_1+\bar{w}_2X_2)-\frac{1}{2}\left(A|w_1|^2+2B\Re(w_1\bar{w}_2)+C|w_2|^2\right)}
\end{equation}
TODO: check that the quadratic form is right.
\begin{equation}
\mathcal{L}_{ML}(x)=\max_{\mathcal{A}} \mathcal{L}(x, \mathcal{A})=\max_{h, w'_1, w'_2} e^{h\Re(\bar{w}'_1 X_1+\bar{w}'_2X_2)-\frac{1}{2}h^2\left(A|w'_1|^2+2B\Re(w'_1\bar{w}'_2)+C|w'_2|^2\right)}
\end{equation}
\begin{equation}
\mathcal{F}(x)=\max_{h, w'_1, w'_2} h\Re(\bar{w}'_1 X_1+\bar{w}'_2X_2)-\frac{1}{2}h^2(A|w'_1|^2+2B\Re(w'_1\bar{w}'_2)+C|w'_2|^2)
\end{equation}
\begin{equation}
\mathcal{F}(x)=\max_{w'_1, w'_2} \frac{\left(\Re(\bar{w}'_1 X_1+\bar{w}'_2X_2)\right)^2}{A|w'_1|^2+2B\Re(w'_1\bar{w}'_2)+C|w'_2|^2}
\end{equation}
If we were to maximize over $w_1$ and $w_2$ right away, we get the following system of equations:
\begin{equation}
\begin{array}{rcl}
\frac{1}{2}X_1&=&\frac{1}{2}\left(Aw_1+Bw_2\right)\\
\frac{1}{2}X_2&=&\frac{1}{2}\left(Bw_1+Cw_2\right)\\
\end{array}
\end{equation}
\begin{equation}
\begin{array}{rcl}
w_1&=&\frac{1}{AC-B^2}\left(CX_1-BX_2\right)\\
w_2&=&\frac{1}{AC-B^2}\left(-BX_1+AX_2\right)\\
\end{array}
\end{equation}
\begin{equation}
\mathcal{F}(x)=\frac{1}{2}\frac{1}{AC-B^2}\left(C|X_1|^2-2B\Re(\bar{X_1}X_2)+A|X_2|^2\right)
\end{equation}
}

\check{
Derivation of $\mathcal B$-statistic:
\begin{equation}
\mathcal{L}(x, \mathcal{A})=e^{\mathcal{A}^\mu x_\mu-\frac{1}{2}\mathcal{A}^\mu M_{\mu\nu} \mathcal{A}^\nu} =e^{\Re(\bar{w}_1 X_1+\bar{w}_2X_2)-\frac{1}{2}\left(A|w_1|^2+2B\Re(w_1\bar{w}_2)+C|w_2|^2\right)}
\end{equation}
TODO: check that the quadratic form is right.
\begin{equation}
\mathcal{B}(x)=\int_{h<h_0} d{\mathcal{A}} \mathcal{L}(x, \mathcal{A})=\int_0^{h_0}dh \int dw' e^{h\Re(\bar{w}'_1 X_1+\bar{w}'_2X_2)-\frac{1}{2}h^2(A|w'_1|^2+2B\Re(w'_1\bar{w}'_2)+C|w'_2|^2)}
\end{equation}
\begin{equation}
\mathcal{B}(x)=\int dw' \int_0^{h_0}dh e^{h\Re(\bar{w}'_1 X_1+\bar{w}'_2X_2)-\frac{1}{2}h^2(A|w'_1|^2+2B\Re(w'_1\bar{w}'_2)+C|w'_2|^2)}
\end{equation}
\begin{equation}
\mathcal{B}(x)=\int dw' \int_0^{h_0}dh e^{h\Re(\bar{w}'_1 X_1+\bar{w}'_2X_2)-\frac{1}{2}h^2(A|w'_1|^2+2B\Re(w'_1\bar{w}'_2)+C|w'_2|^2)}
\end{equation}
\begin{equation}
\mathcal{B}(x)=\int dw' \int_0^{h_0}dh e^{h\Re(\bar{w}'_1 X_1+\bar{w}'_2X_2)-\frac{1}{2}h^2(A|w'_1|^2+2B\Re(w'_1\bar{w}'_2)+C|w'_2|^2)}
\end{equation}
\begin{equation}
\int_0^\infty dx e^{ax-bx^2/2}=e^{\frac{a^2}{2b}} \int_0^\infty dx e^{-b(x-a/b)^2/2}=\frac{1}{\sqrt{b}}e^{\frac{a^2}{2b}} \int_0^\infty dx e^{-(x-a/\sqrt{b})^2/2}
\end{equation}
\begin{equation}
\int_0^\infty dx e^{ax-bx^2/2}=\frac{1}{\sqrt{b}}e^{\frac{a^2}{2b}} \int_{-a/\sqrt{b}}^\infty dx e^{-x^2/2}=\frac{1}{\sqrt{b}}e^{\frac{a^2}{2b}} \int_{-\infty}^{a/\sqrt{b}} dx e^{-x^2/2}
\end{equation}
\begin{equation}
\mathcal{B}(x)=\int dw' \frac{e^{\SNR_R(w_1', w_2')/2}}{\sqrt{\mathfrak{A}(w_1', w_2')}}\int_{-\infty}^{\SNR_R(w_1', w_2')/2}e^{-x^2/2}dx
\end{equation}
\begin{equation}
\mathcal{B}(x)=\int_{\phi=0} dw' \frac{1}{2\pi}\int_0^{2\pi} d\phi \frac{e^{\cos(\phi)\SNR(w_1', w_2')/2}}{\sqrt{\mathfrak{A}(w_1', w_2')}}\int_{-\infty}^{\cos(\phi)\SNR(w_1', w_2')/2}e^{-x^2/2}dx
\end{equation}
To proceed further we need efficient means of computing
$$
\Theta(x)=\frac{1}{2\pi}\int_0^{2\pi} \frac{e^{\cos(\phi)x}}{\sqrt{2\pi}}\int_{-\infty}^{\cos(\phi)x}e^{-s^2/2}ds d\phi 
$$
We have the following easy identities:
$$
\Theta(x)=\Theta(-x)
$$
$$
\Theta(0)=\frac{1}{2}
$$
It is also easy to compute approximations for small and large $x$:
$$
\Theta(x)=\frac{1}{2}+\left(\frac{1}{8}+\frac{1}{2\sqrt{2\pi}}\right)x^2+O(x^4)
$$
$$
\Theta(x)=\frac{e^{|x|}}{\sqrt{2\pi |x|}}(1+O(1/x))
$$
%
%
Numeric approximation valid for all $x$ to $0.3$\%:
$$
\Theta(x)=\frac{\exp\left({\sqrt{a+x^2}+b_1+b_2\cos\left(\pi\frac{x}{\sqrt{x^2+c^2}+c}\right)+b_3\cos\left(3\pi\frac{x}{\sqrt{x^2+c^2}+c}\right)}\right)}{\left(4\pi^2x^2+16e^{4\sqrt{a}}\right)^{1/4}}
$$
Better approximation:
$$
\Theta(x)=\frac{\exp\left({\sqrt{0.25+x^2}}\right)}{\left(4\pi^2x^2+16e^{2}\right)^{1/4}}\frac{a_0+a_2x^2+a_4x^4+x^6}{b_0+b_2x^2+b_4x^4+x^6}
$$
has $0.05$\% error over the entire range with the following values of constants:
$$
\begin{array}{l}
a_0=7.71990148904874\\
a_2=19.0337266315871\\
a_4=5.20172247607555\\
b_0=7.7201854234519\\
b_2=21.1533518190664\\
b_4=4.28188537828521 \\
\end{array}
$$
}

\check{
Approximation of normal distribution:
\begin{equation}
\frac{1}{\sqrt{2\pi}}\int_{-\infty}^x e^{-s^2/2}ds=1/(1+\exp(-x(1.6+0.07x^2-0.0005x^4)))
\end{equation}
Good to within $0.038$\%.
Approximation of exponent times normal distribution:
$$
e^x\frac{1}{\sqrt{2\pi}}\int_{-\infty}^x e^{-s^2/2}ds=e^x-0.525e^{-0.3(x-0.25)^2}
$$
Good to within $4.2$\%.
\\
}

There is an easy and elegant way to compute all of these statistics with minimal cost. 

First we note, that our statistics are monotonic in signal strength; they just disagree as to which signal parameters are given preference. Once the signal strength is fixed (in any statistic), though, the rest of the parameters form a bounded manifold - and other statistics achieve a maximum and minimum value on it. We can thus infer an estimate of another statistic from knowing the maximum of some convenient, easy-to-compute measure of signal strength. 

As a toy example, assume that our statistics vary by at most a factor of $4$ for signals of the same power $|X_1|^2+|X_2|^2$ and that our signal array consists of $\approx~300000$ elements. The average maximum of $300000$ of mean=$1$, exponentially distributed samples is $12$. 
But $95$\% of these numbers are below $3=12/4$. Thus, to find the maximum of a more complicated statistics we only need to examine  $5$\% of the samples, providing a factor of $20$ speed up. If a signal is present our maximum is even higher excluding a larger amount of data.

\begin{figure}[!htbp]
\begin{center}
  \includegraphics[width=3.5in]{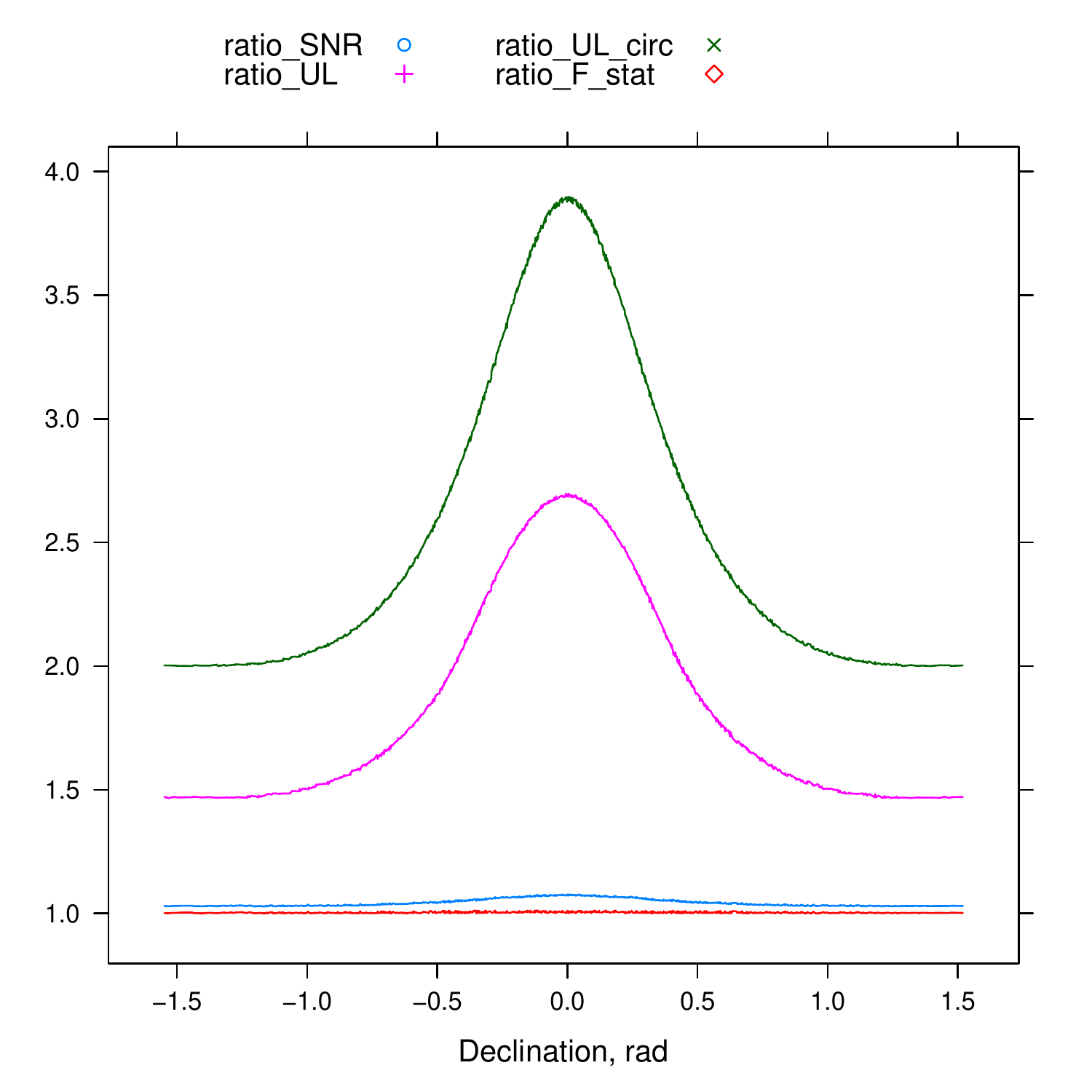}

 \caption{Maximum variance of statistics for signals of constant adjusted power. The numbers come from simulations using Gaussian noise and describe the ratio between the maximum and minimum statistic values for signal of constant adjusted power. Top curve - upper limit assuming circular polarization. Next curve below is upper limit statistic, followed by $\SNR$ and $\mathfrak F$-statistic (color online). }
\label{fig:estimate_ratios}
\end{center}
\end{figure}

\begin{figure}[!htbp]
\begin{center}
  \includegraphics[width=3.5in]{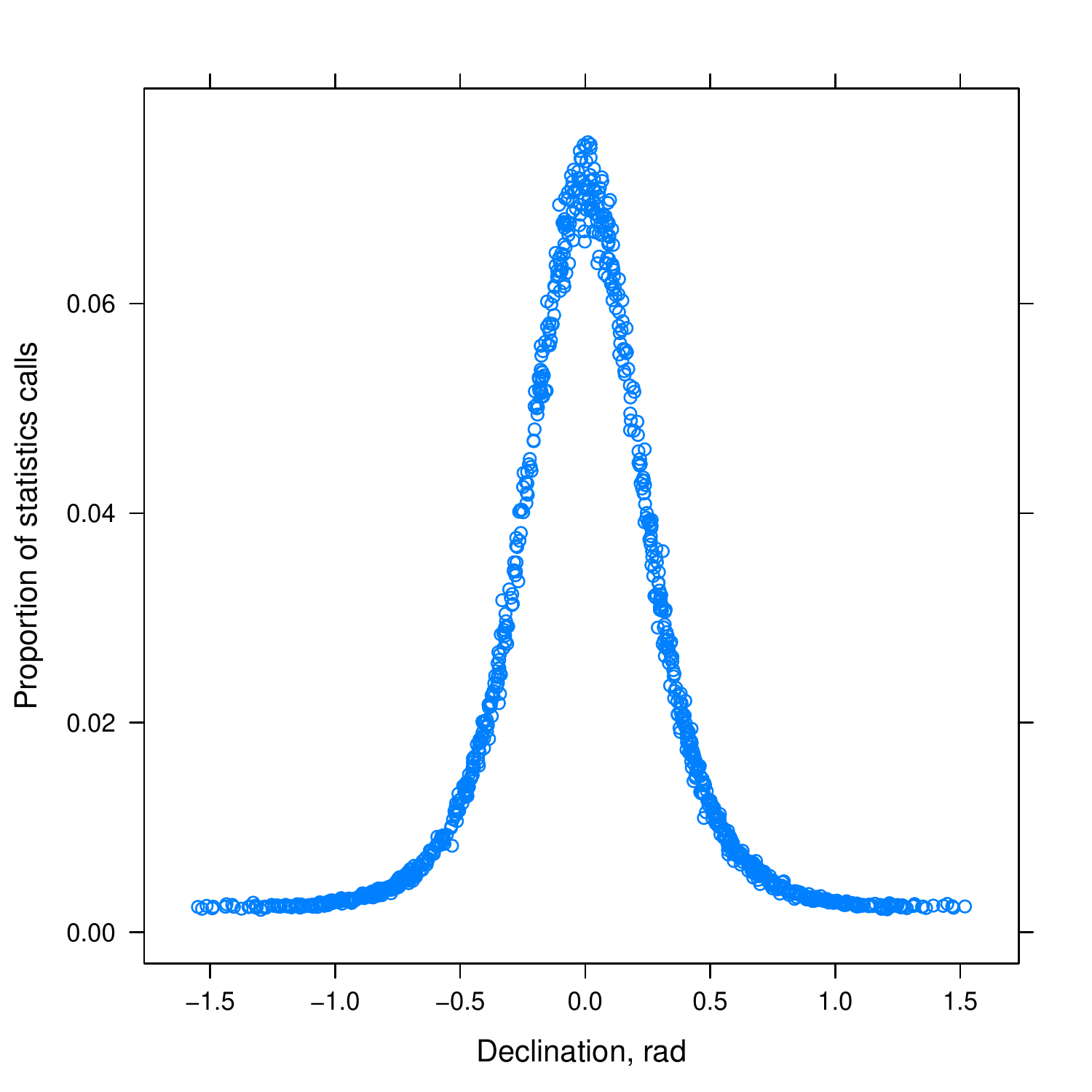}

 \caption{Fraction of templates resulting in statistic calls. The underlying data was pure Gaussian noise (color online).}
\label{fig:statistics_fraction}
\end{center}
\end{figure}

In a practical implementation it is better to use adjusted power (see table \ref{tab:statistics}), which results in a less-than-$4$ worst-case scale factor for upper limit statistics (evaluated for strong signals) and close to unity factors for plain SNR and $\mathfrak F$-statistics as can be seen on figure \ref{fig:estimate_ratios}. The strong dependence on declination is due to the antenna pattern of the detector. The fraction of templates for which we computed upper limit statistic during a Gaussian noise simulation run (discussed in more detail in the following section) is shown in figure \ref{fig:statistics_fraction}.

\section{\label{sec:simulations}Performance and validation}

An initial implementation of the ideas discussed above has been completed. It has a number of simplifications compared to an eventual production program - the input data is assumed contiguous, one spindown is analyzed at a time, and there is no support for higher-order source frequency evolution parameters. 

\begin{figure}[!htbp]
\begin{center}
  \includegraphics[width=3.5in]{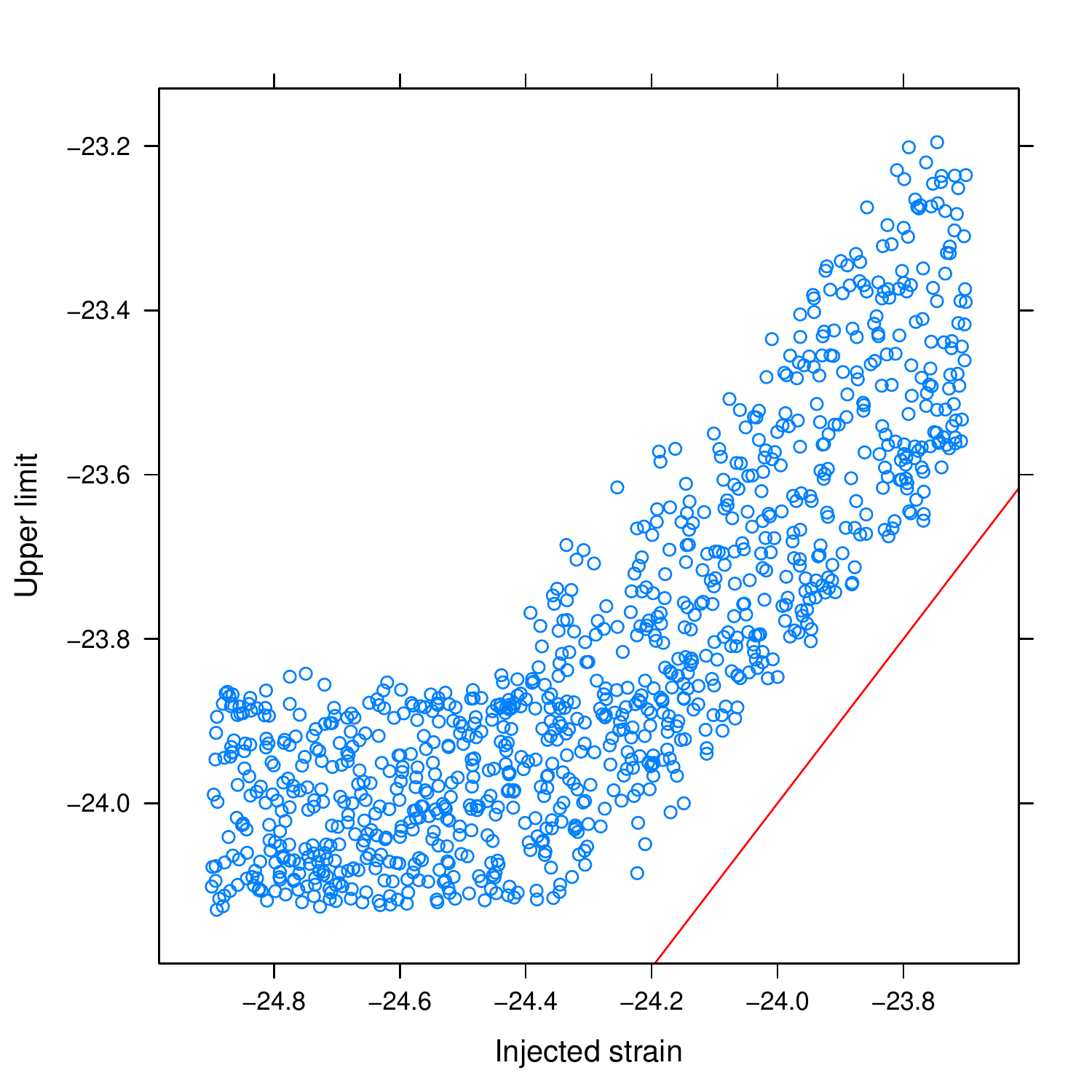}

 \caption{Upper limit versus injected strain (color online).}
\label{fig:ul_vs_strain}
\end{center}
\end{figure}

We have performed Monte-Carlo runs of 1000 injections each into Gaussian data spanning four million seconds (approximately $1.5$ months), assuming a detector located at LIGO Hanford observatory. The injection sky locations and source orientation were uniformly distributed. The spindown parameter was set to be zero. In the first run, we injected signals of various strength (figure \ref{fig:ul_vs_strain}) to test signal detection and upper limit estimation. The second run had identical parameters and noise distribution, but the signal strengths were set to $0$. This provided upper limits on pure noise alone (for relative comparison) as well as worst-case computational performance, as the presence of strong signals makes computation of statistics faster. The injection frequency was uniformly distributed in  $\pm 0.084$~Hz interval centered on $400$~Hz.

Each separate injection run included a search over a $1$-arcminute disk around a nominal sky location that was obtained from rounding the true injection locations; hence the actual injection locations were uniformly distributed in relation to the sampled grid.

The upper limit plot in Figure \ref{fig:ul_vs_strain} shows that upper limits are consistently above the injected signal values. A gap between the reconstructed points and the red line marking injected values is due to a conservative correction for Hann-windowed input data.

\begin{figure}[!htbp]
\begin{center}
  \includegraphics[width=3.5in]{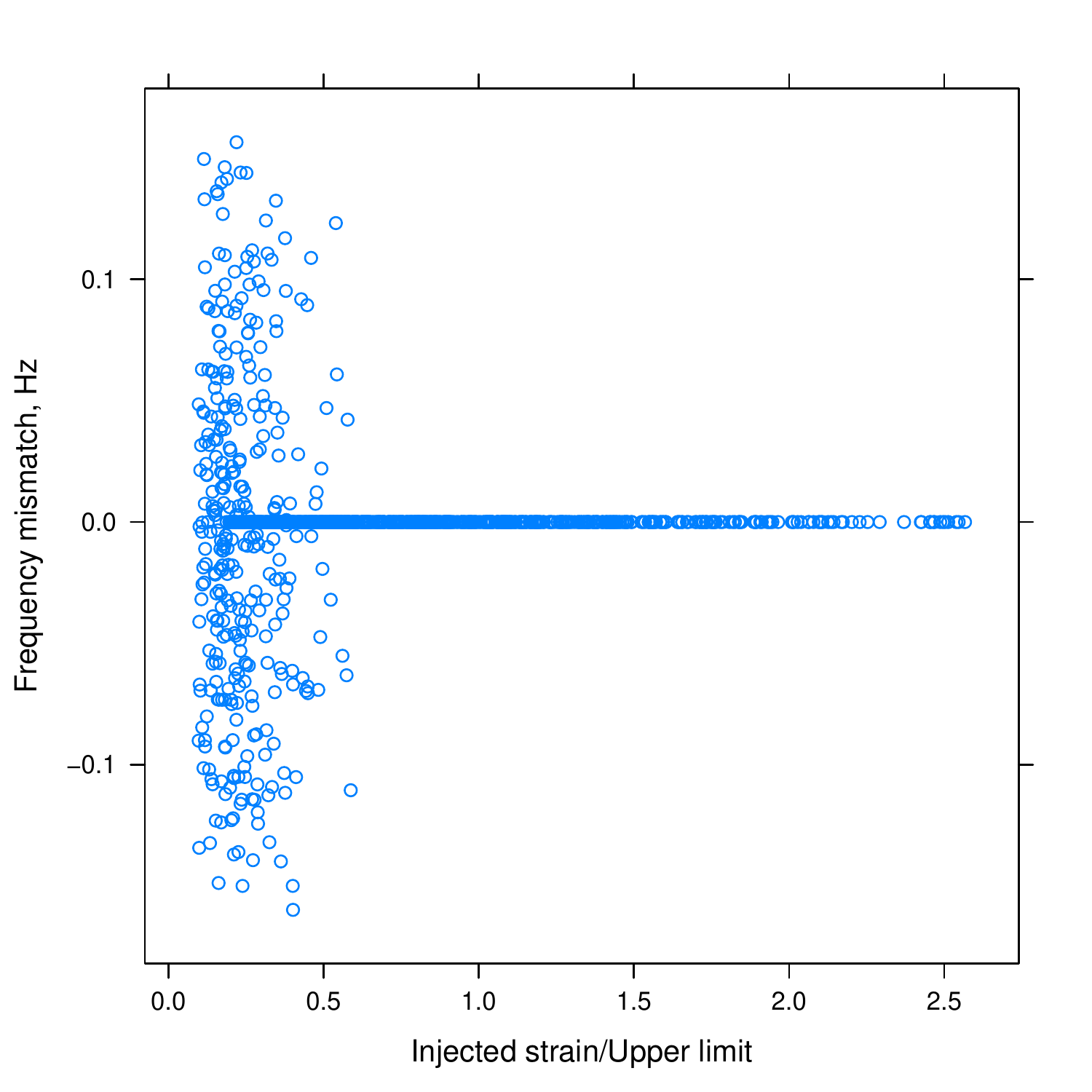}

 \caption{Frequency reconstruction (color online).}
\label{fig:relative_f0_reconstruction}
\end{center}
\end{figure}

When signals rise above background their frequencies are well localized (figure \ref{fig:relative_f0_reconstruction}). We use this localization as a criterion for detection: a signal is considered to be found if its frequency is within $\sci{1}{-5}$~Hz of true value. This corresponds to false alarm ratio of $\approx\sci{6}{-5}$. 

\begin{figure}[!htbp]
\begin{center}
  \includegraphics[width=3.5in]{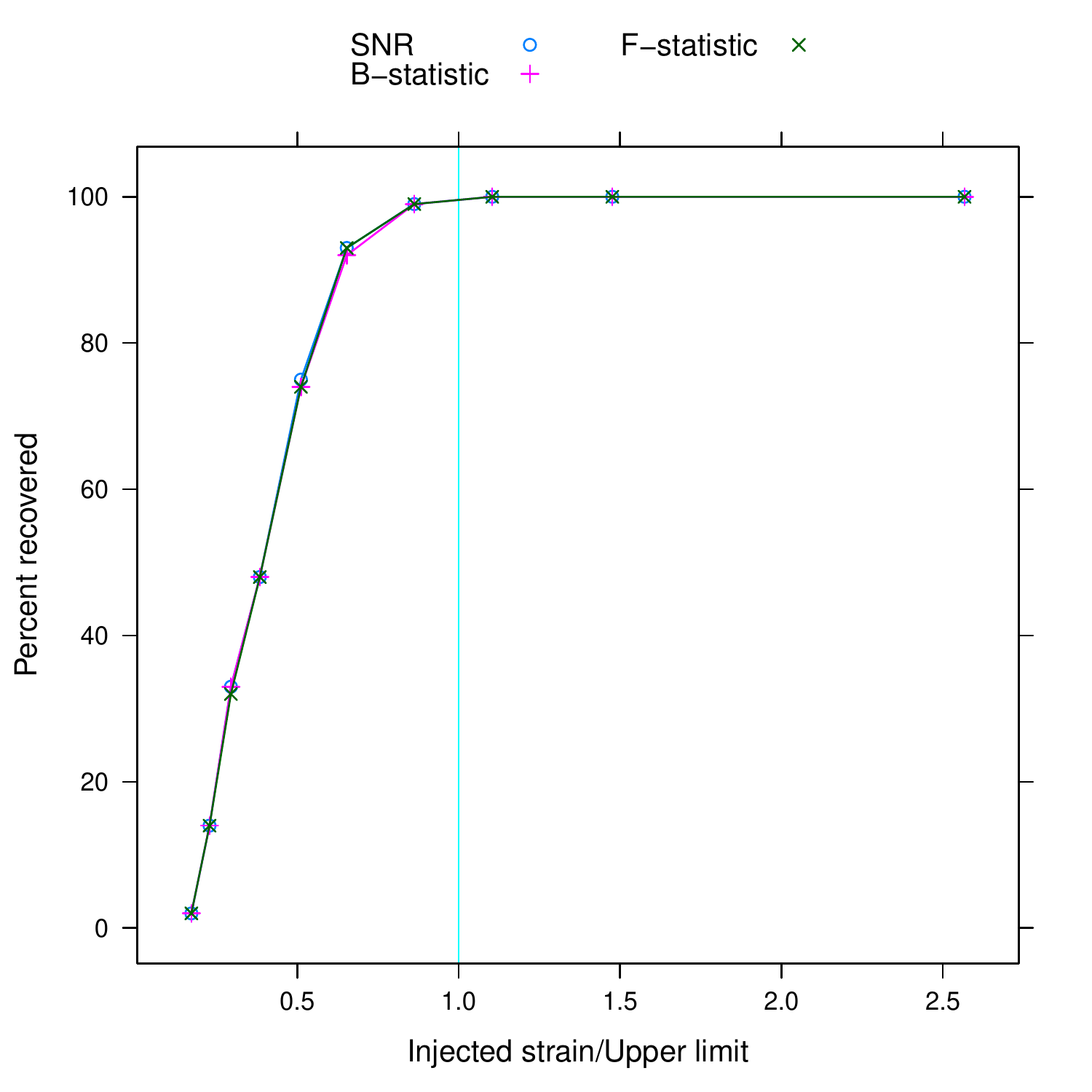}

 \caption{Statistics efficiency versus strength of injected signal, relative to established upper limit (color online).}
\label{fig:relative_efficiency_vs_strain}
\end{center}
\end{figure}

Figure \ref{fig:relative_efficiency_vs_strain} compares the efficiencies of various detection statistics. The SNR and $\mathfrak F$-statistics are mathematically equivalent; the only difference is that the SNR is computed by iterating over a grid of parameters $w_1'$ and $w_2'$, while the $\mathfrak F$-statistic has a much more efficient closed form. The slower SNR algorithm was used as a bridge to an implementation of the $\mathfrak B$-statistic, and will also be useful for future implementation of $\delta>0$ loosely coherent statistics.

As seen from the plot we observe no difference in performance between $\mathfrak F$-statistic and $\mathfrak B$-statistic. We believe this is due to two factors:
\begin{itemize}
 \item First, even for pure noise, maximizing over $\sci{2}{6}$ bins results in high $\SNR$ values where the difference between $\mathfrak F$-statistic and $\mathfrak B$-statistic is smaller.
 \item Secondly, the area searched is small which significantly reduces influence of the weight term $\mathfrak A(w_1', w_2')$ that appears in the $\mathfrak B$-statistic
\end{itemize}
It might be possible to take advantage of the improved performance of the $\mathfrak B$-statistic at low $\SNR$ by analyzing multi-detector data with consistency cuts to bring down maximum $\SNR$.

\begin{figure}[!htbp]
\begin{center}
  \includegraphics[width=3.5in]{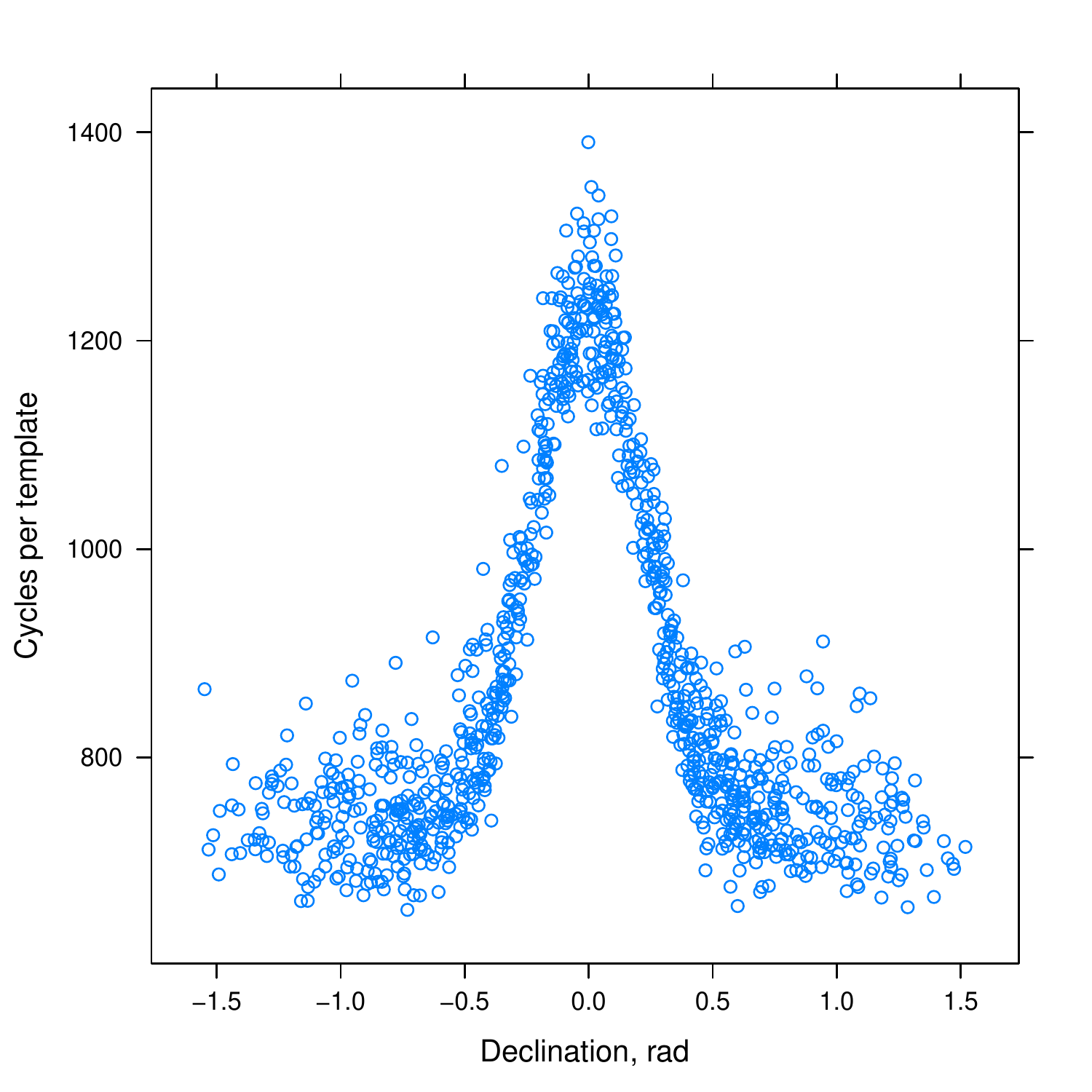}

 \caption{CPU cycles spent per template as a function of declination of the injected signal (color online).}
\label{fig:template_cycles_vs_dec}
\end{center}
\end{figure}

The computational performance of our code is shown in Figure \ref{fig:template_cycles_vs_dec}. The $y$-axis is in units of cycles per template, with each template computed once per frequency, sky position and spindown while sampling all possible alignments of the source. All statistics ($\SNR$, upper limit, circular upper limit, $\mathfrak F$-statistic and $\mathfrak B$-statistic) were computed during the run. The underlying data was purely Gaussian. The simulations were run on a cluster of $2.3$~GHz AMD processors.

Our worst-case performance is below $1500$ cycles which favorably compares with performance of resampling \cite{resampling,joe} that is estimated to be $\approx 20000$ cycles per template. The cpu utilization by different parts of the algorithms is shown in table \ref{tab:cpu_cycles_per_function}. Only a third of the cycles is attributable to computation of the convolution, while at least 46\% is spent gathering statistics, leaving much room for further improvement.

\begin{table}[!htbp]
\begin{center}
\begin{tabular}{r|l}
CPU fraction & Code description \\ 
\hline \hline  
36\% &  Computation of upper limit \\
16\% &  Computation of 11-term convolution \\
10\% &  Computation of terms of convolutions \\
8\%  &  General statistics function \\
7\% & Computation of 63-term pseudo-convolution \\
2\% & Computation of logarithm \\
Balance of 21\% & Spread out over many parts of the algorithm\\
 
\end{tabular}
 \caption{CPU cycles spent in different parts of the algorithm for sky position with declination of $1.0$ radian.}
\label{tab:cpu_cycles_per_function}
\end{center}
\end{table}

\section{Conclusions}
We have described an implementation of the loosely coherent statistic that searches over families of ideal continuous gravitational signals.
The performance of the algorithm is more than an order of magnitude better than previously published algorithms, opening the way for exploring wider parameter spaces.

There are several directions for further improvement:
\begin{itemize}
 \item Making use of coherent or loosely coherent combination of data from two interferometers should improve sensitivity and provide better rejection of detector artifacts. This might also be an area where the $\mathfrak B$-statistic will show its strength.
 \item It is necessary to derive more efficient alternatives for the computation of upper limit.
\item It would be desirable to extend the algorithm to search over frequency evolution parameters to be able to detect binary systems with weak modulation.
\item The fine granularity of input data can be used to avoid high-intensity glitches, by excluding contaminated SFTs.
\end{itemize}

\section{Acknowledgments}
This work has been done while being a member of LIGO laboratory, supported by funding from United States National Science Foundation.
The author would like to thank Bruce Allen for the invaluable opportunity to stay at Max Planck Institute for Gravitational Physics (Albert Einstein Institute).

 The author has greatly benefited
from suggestions and comments of his colleagues, in particular Reinhard Prix and Keith Riles.

This document has been assigned LIGO Laboratory document number \texttt{LIGO-P1100104-v6}.

\appendix

\section{Efficient computation of $\mathfrak B$-statistic}
$\mathfrak B$-statistic was introduced in \cite{BayesianFstat} as a Bayesian alternative to $\mathfrak F$-statistic. It was shown that the $\mathfrak F$-statistic is equivalent to a Bayesian statistic with a prior that favors linearly polarized signals and high signal strength. In contrast, the $\mathfrak B$-statistic is isotropic in spin orientation.

The statistic starts with the likelihood function:
\begin{equation}
\begin{array}{l}
\mathcal{L}(x, \mathcal{A})=e^{\mathcal{A}^\mu x_\mu-\frac{1}{2}\mathcal{A}^\mu M_{\mu\nu} \mathcal{A}^\nu} =\\
\quad=e^{\Re(\bar{w}_1 X_1+\bar{w}_2X_2)-\frac{1}{2}\left(Y_{11}|w_1|^2+2Y_{12}\Re(w_1\bar{w}_2)+Y_{22}|w_2|^2\right)}
\end{array}
\end{equation}
Instead of maximizing it, which is the approach of the $\mathfrak F$-statistic we compute the integral:
\begin{equation}
\begin{array}{l}
\mathcal{B}(x)=\displaystyle\int_{h<h_0} d{\mathcal{A}} \mathcal{L}(x, \mathcal{A})=\displaystyle \int dw' \int_0^{h_0}dh \cdot\\
\quad \cdot e^{h\Re(\bar{w}'_1 X_1+\bar{w}'_2X_2)-\frac{1}{2}h^2(Y_{11}|w'_1|^2+2Y_{12}\Re(w'_1\bar{w}'_2)+Y_{22}|w'_2|^2)}
\end{array}
\end{equation}
The measures $d\mathcal A$ and $dw'$ are chosen to be uniform in parameters $\phi$, $\psi$ and $\cos(\iota)$. The integral with respect to $h$ is not normalized, which allows to set its upper limit infinite. This effectively changes $B(x)$ to be in units of strain. It would be interesting to explore the possibility of deriving an upper limit estimator based on the same principles as $B(x)$.

The integral with respect to $h$ can be shown to be a function of $\textrm{SNR}_R(w_1', w_2')$ and total weight $\mathfrak A(w_1', w_2')$:
\begin{equation}
\int_0^\infty dx e^{ax-bx^2/2}=\frac{1}{\sqrt{b}}e^{\frac{a^2}{2b}} \int_{-\infty}^{a/\sqrt{b}} dx e^{-x^2/2}
\end{equation}
\begin{equation}
\mathcal{B}(x)=\int dw' \frac{e^{\SNR_R(w_1', w_2')/2}}{\sqrt{\mathfrak{A}(w_1', w_2')}}\int_{-\infty}^{\SNR_R(w_1', w_2')/2}e^{-x^2/2}dx
\end{equation}
This still leaves a three dimensional integral to carry out which is undesirable inside the inner loop. We note that $\mathfrak A(w_1', w_2')$ does not depend on phase $\phi$, while 
\begin{equation}
\textrm{SNR}_R(w_1', w_2')=\cos(\phi)\textrm{SNR}(w_1', w_2')
\end{equation}
We can thus represent our statistic as
\begin{equation}
\mathcal{B}(x)=\int_{\phi=0} dw' \frac{1}{\sqrt{\mathfrak{A}(w_1', w_2')}}\Theta\left(\frac{1}{2}\SNR(w_1', w_2')\right)
\end{equation}
where $\Theta(x)$ is defined as
\begin{equation}
\Theta(x)=\frac{1}{2\pi}\int_0^{2\pi} \frac{e^{\cos(\phi)x}}{\sqrt{2\pi}}\int_{-\infty}^{\cos(\phi)x}e^{-s^2/2}ds d\phi 
\end{equation}
We can now study $\Theta(x)$ as a new special function and find a means to compute it efficiently.
We have the following easy identities:
\begin{equation}
\begin{array}{rcl}
\Theta(x)&=&\Theta(-x)\\
\Theta(0)&=&\frac{1}{2}
\end{array}
\end{equation}
It is also easy to compute approximations for small and large $x$:
\begin{equation}
\begin{array}{rcl}
\Theta(x)&=&\frac{1}{2}+\left(\frac{1}{8}+\frac{1}{2\sqrt{2\pi}}\right)x^2+O(x^4)\\
\Theta(x)&=&\frac{e^{|x|}}{\sqrt{2\pi |x|}}(1+O(1/x))
\end{array} 
\end{equation}
Armed with these relations, we can spend some time in numerical experimentation and arrive at the following approximation:
\begin{equation}
\Theta(x)\approx\frac{\exp\left({\sqrt{0.25+x^2}}\right)}{\left(4\pi^2x^2+16e^{2}\right)^{1/4}}\frac{a_0+a_2x^2+a_4x^4+x^6}{b_0+b_2x^2+b_4x^4+x^6}
\end{equation}
which has a $0.05$\% error over the entire range with the following values of constants:
\begin{equation}
\begin{array}{rcD{.}{.}{10}}
a_0&=&7.7199014890487\\
a_2&=&19.0337266315871\\
a_4&=&5.2017224760755\\
b_0&=&7.7201854234519\\
b_2&=&21.1533518190664\\
b_4&=& 4.2818853782852 \\
\end{array}
\end{equation}
Now we can compute $\mathfrak B$-statistic with a simple sum over a uniform grid in $\psi$ and $\cos(\iota)$, at a cost within an order of magnitude of computing the $\SNR$ statistic.


\end{document}